\documentclass[prl,aps,showpacs]{revtex4}
\usepackage{graphicx}

\begin{document}

\title{Influence of Zeeman splitting and thermally excited polaron
states on magneto-electrical and magneto-thermal properties of
magnetoresistive polycrystalline manganite
$La_{0.8}Sr_{0.2}MnO_3$}
\author{S. Sergeenkov}
\affiliation{Departamento de F\'isica, CCEN, Universidade Federal
da Para\'iba, Cidade Universit\'aria, 58051-970 Jo\~ao Pessoa, PB,
Brazil}
\author{J. Mucha}
\affiliation{Institute of Low Temperature and Structure Research,
Polish Academy of Sciences, P.O. Box 1410, 50-950 Wroclaw, Poland}
\author{M. Pekala}
\affiliation{Department of Chemistry, University of Warsaw, Al.
Zwirki i Wigury 101, PL-02-089 Warsaw, Poland}
\author{V. Drozd}
\affiliation{Department of Chemistry, Kiev National Taras
Shevchenko University, 60 Volodymyrska st, Kiev, 01033 Ukraine and
Center for Study Matter at Extreme Conditions, Florida
International University, Miami, USA}
\author{M. Ausloos}
\affiliation{SUPRATECS, Institute of Physics B5, University of
Li$\grave e$ge, B-4000 Li$\grave e$ge, Belgium}

\date{\today; accepted for publication in Journal of Applied Physics}

\begin{abstract}
Some possible connection between spin and charge degrees of
freedom in magneto-resistive manganites is investigated through a
thorough experimental study of the magnetic (AC susceptibility and
DC magnetization) and transport (resistivity and thermal
conductivity) properties. Measurements are reported in the case of
well characterized polycrystalline $La_{0.8}Sr_{0.2}MnO_3$
samples. The experimental results suggest rather strong
field-induced polarization effects in our material, clearly
indicating the presence of ordered FM regions inside the
semiconducting phase. Using an analytical expression which fits
the spontaneous DC magnetization, the temperature and magnetic
field dependences of both electrical resistivity and thermal
conductivity data are found to be well reproduced through a
universal scenario based on two mechanisms: (i) a magnetization
dependent spin polaron hopping influenced by a Zeeman splitting
effect, and (ii) properly defined thermally excited polaron states
which have to be taken into account in order to correctly describe
the behavior of the less conducting region. Using the
experimentally found values of the magnetic and electron
localization temperatures, we obtain $L=0.5nm$ and $m_p=3.2m_e$
for estimates of the localization length (size of the spin
polaron) and effective polaron mass, respectively.
\end{abstract}


\maketitle

\section{I. Introduction}

Manganite systems exhibit a rich collection of interesting and
intriguing properties, which can be tailored for a wide variety of
applications (such as low-loss power delivery, quantum computing,
ultra high-density magnetic data storage and more recently
spintronic applications). Many such oxides have been prepared in
bulk form or as thin films, which paved the way for intensive
research studies in the past several decades (see, e.g.,
~\cite{1,2,3,4,5,6,7} and further references therein). Among many
different complex manganese oxides the most actively studied are
the $(R_{1-x}A_x)MnO_3$ series where $R(A)$ stands for a trivalent
rare-earth (divalent alkaline-earth) cation.

It was the discovery of the so-called colossal magnetoresistance
(CMR) in these materials that spurred their comprehensive and
thorough study. Soon enough it was realized that CMR exhibiting
materials possess many features which make them very fascinating.
First of all, these compounds are found to undergo a distinctive
double phase transition under cooling from a paramagnetic (PM)
weakly conductive insulating-like  (I) state to a ferromagnetic
(FM) more conductive metallic-like (M) state. These two
transitions occur at what is conveniently described as the Curie
temperature $T_C$ and the so-called charge carrier localization
temperature $T_{MI}$, respectively. The observable difference
between the two critical temperature values is usually attributed
to the quality of the sample. However, even for perfect
(defect-free) single crystals these two temperatures are not
exactly equal, due to inseparable correlations between charge and
spin degrees of freedom which are admitted to be the causes behind
the observable CMR phenomena. Besides, in real materials these
interactions are always modified by both intrinsic and extrinsic
inhomogeneities (for example, the parameters of both transitions
are known to be very sensitive to the oxygen content).

Since no comprehensive theory which would explain {\it all} the
complexity of this interesting phenomenon has been suggested so
far, it is still very important to extract microscopic parameters
from real measurements and compare them to theoretical forecasts.
Many routes can be used to investigate or sort out the (likely)
numerous (though basic) underlying mechanisms. It is now well
established that the complicated phase diagram of
magneto-resistive (MR) manganites is very sensitive to $Mn$ site
substitution and to magnetic fields. Beyond the usually admitted
primo scenarios (in terms of the double exchange mechanism), the
structure sensitive Jahn-Teller effect and the strong
electron-phonon coupling are found to play an important role in
these materials~\cite{2}. Besides, in the low temperature
conducting ferromagnetic phase, clear evidence for a collective
magnon signature (in the form of the  $T^{3/2}$ Bloch law) was
found and attributed to the so-called magnon-polaron
excitations~\cite{3}. It was also pointed out that while some
features are better explained through localized (spin) states
alone, others definitely require the presence of collective
excitations (or both) for their explanation~\cite{1,2,3}. It would
be also interesting to have complementary information, both away
from and including the transition regions.

Some possible connection between spin and charge degrees of
freedom in magneto-resistive manganites is sometimes investigated
through either experimental studies of the magnetic properties or
through transport properties. However for better understanding of
the underlying physical mechanisms behind CMR like phenomena, it
is always very important to study various properties quasi
simultaneously. We strongly believe that experimental and
theoretical results should corroborate and complement each other.
That is why acquisition of fine reliable data (especially in the
presence of strong magnetic fields) could help verify the modern
theoretical concepts and allow to extract  the values of important
physical parameters with high precision.  Moreover, in view of the
intricate character of the interaction mechanisms involved, it is
quite evident that some features will better manifest themselves
via magneto-electric transport properties while the others will
require more sophisticated magneto-thermal transport measurements.

In the present paper we endeavor to elucidate the field-induced
charge-spin correlations in CMR exhibiting $La_{0.8}Sr_{0.2}MnO_3$
manganites previously studied by many researchers (a nonexhaustive
list can be found in~\cite{4}) through a thorough study of
possible connections between their (equilibrium) magnetic and
(non-equilibrium) transport properties. In achieving this goal, we
have found new and somewhat unexpected results which, in our
opinion, can shed more light on the nature of these interesting
materials. The paper is organized as follows. In Section II we
present the experimental results for our polycrystalline
$La_{0.8}Sr_{0.2}MnO_3$ samples which include both magnetic (AC
susceptibility and DC magnetization) and transport (resistivity
and thermal conductivity) measurements in applied magnetic fields.
A detailed theoretical discussion of the obtained results is given
in Section III. To explain our findings, a universal and coherent
scenario will be put forward based on various components,
including (i) a magnetization dependent spin polaron hopping, (ii)
Zeeman splitting effects, (iii) properly defined thermally excited
polaron states (needed to correctly describe the temperature
behavior of the less conducting region), and (iv) validity of the
Wiedemann-Franz law.  The paper is concluded with a short summary
of the obtained results in Section IV.

\section{II. Experimental results}

\begin{figure*}
\includegraphics[width=7cm,angle=90]{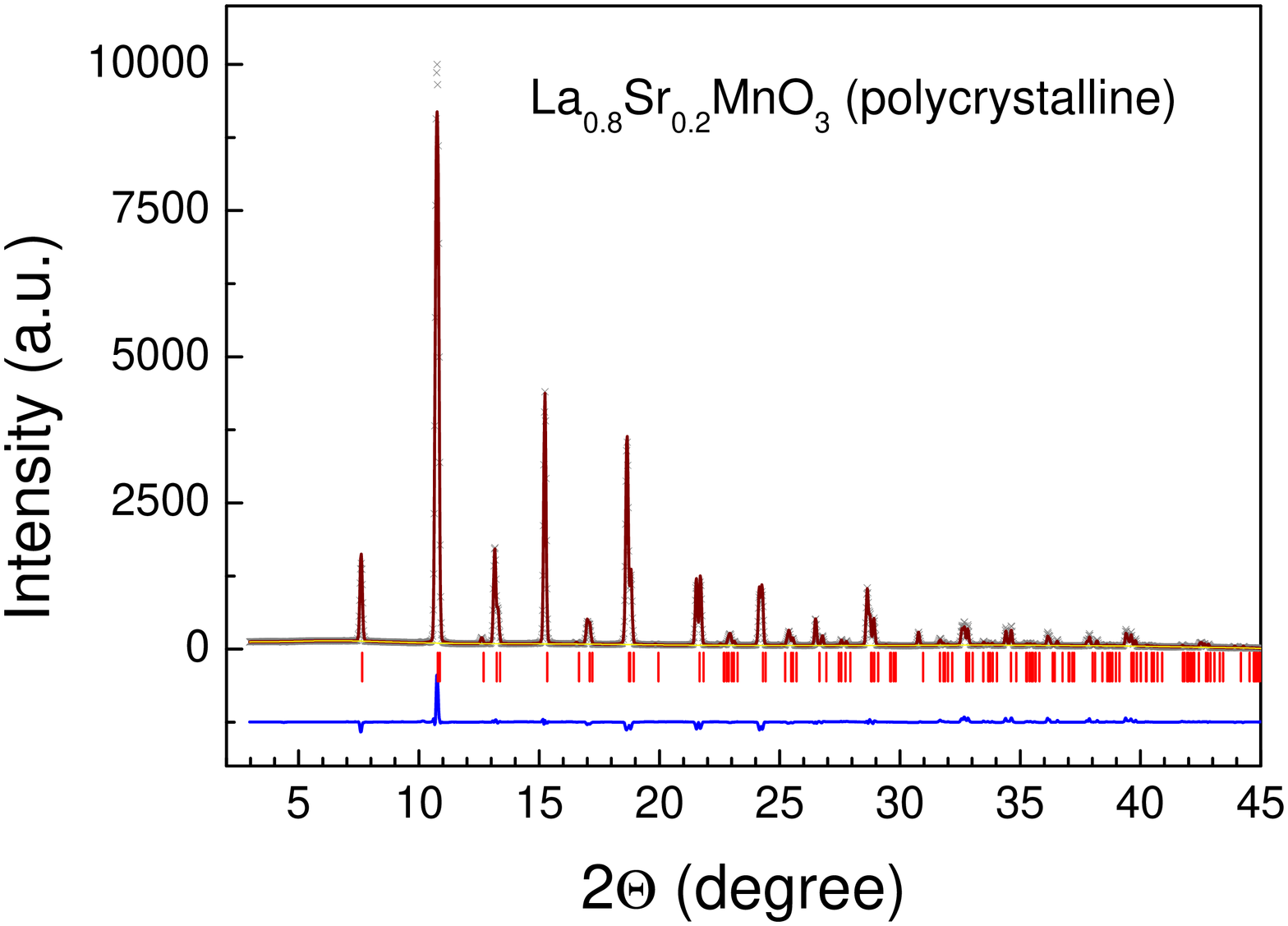}
\caption{\label{fig:1} (Color online) Rietveld refinement plot for
polycrystalline manganite of $La_{0.8}Sr_{0.2}MnO_3$: observed
intensities (symbols: crosses), calculated intensities (wine
line), background (yellow solid line),  reflections (red vertical
lines), and difference (blue solid line at the bottom). }
\end{figure*}

Polycrystalline $La_{0.8}Sr_{0.2}MnO_3$ samples were synthesized
by a carbonate precursor method~\cite{5}. The structural quality
of our samples was verified through X-ray diffraction. Structural
refinements made with X-ray data show (see Fig.~\ref{fig:1}) that
the samples are single phased and rhombohedral with structural
parameters very close to the standard ones~\cite{6,7} (with the
oxygen deficiency less than $0.01$).

\begin{figure*}
\includegraphics[width=7.0cm]{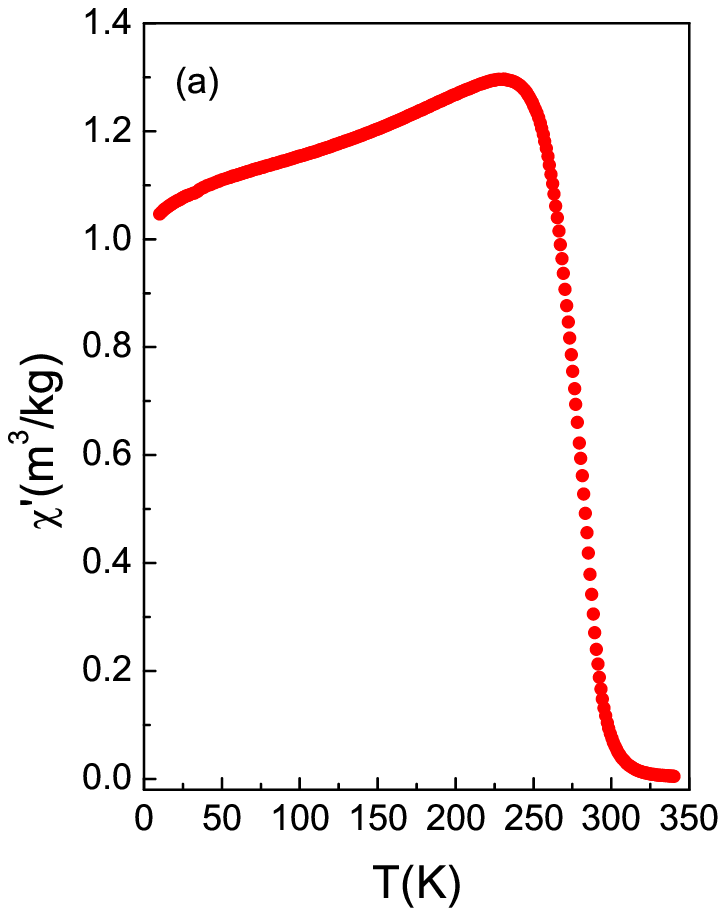}
\includegraphics[width=7.50cm]{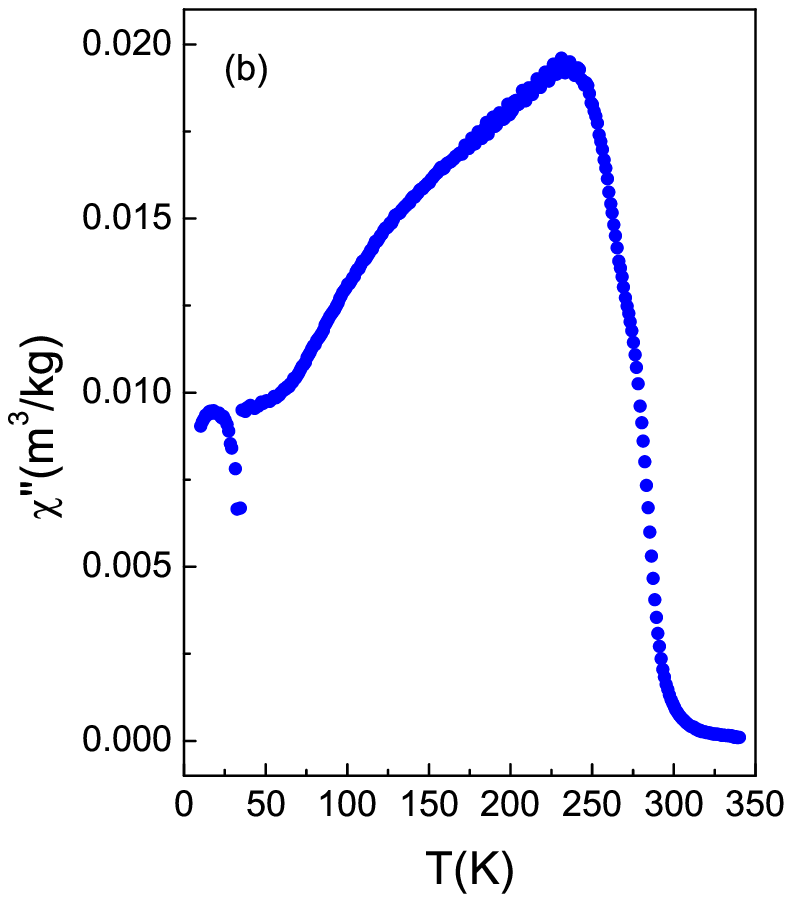}
\caption{\label{fig:2} (Color online) Temperature dependence of
(a) in-phase $\chi '$ and (b) out-of-phase $\chi ''$ components of
the AC magnetic susceptibility of polycrystalline
$La_{0.8}Sr_{0.2}MnO_3$.}
\end{figure*}

\begin{figure*}
\includegraphics[width=7.0cm]{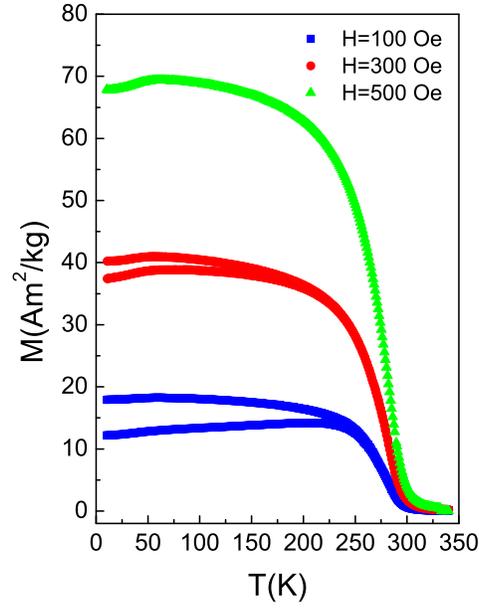}
\caption{\label{fig:3} (Color online) Temperature dependence of DC
magnetization of polycrystalline $La_{0.8}Sr_{0.2}MnO_3$ for
various magnetic fields. The lower and upper branches correspond
to zero-field-cooled (ZFC) and field-cooled (FC) curves,
respectively.}
\end{figure*}

\begin{figure*}
\includegraphics[width=8cm]{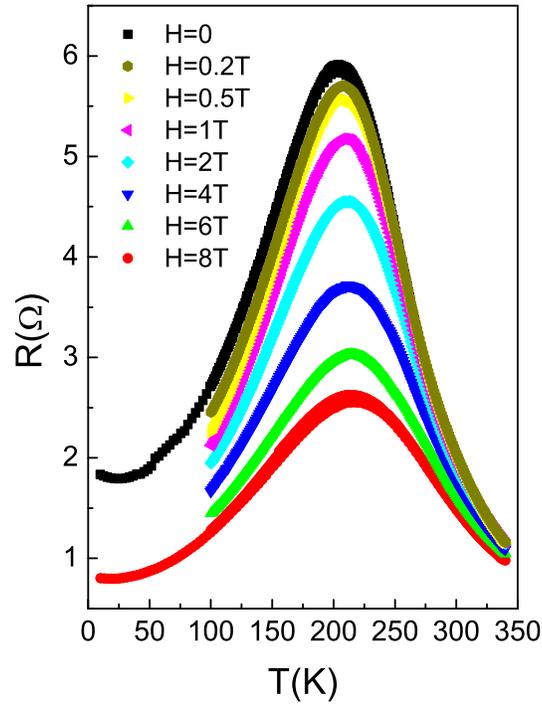}
\caption{\label{fig:4} (Color online) Temperature dependence of
electrical resistance of polycrystalline $La_{0.8}Sr_{0.2}MnO_3$
for various magnetic fields.}
\end{figure*}

\begin{figure*}
\includegraphics[width=8cm]{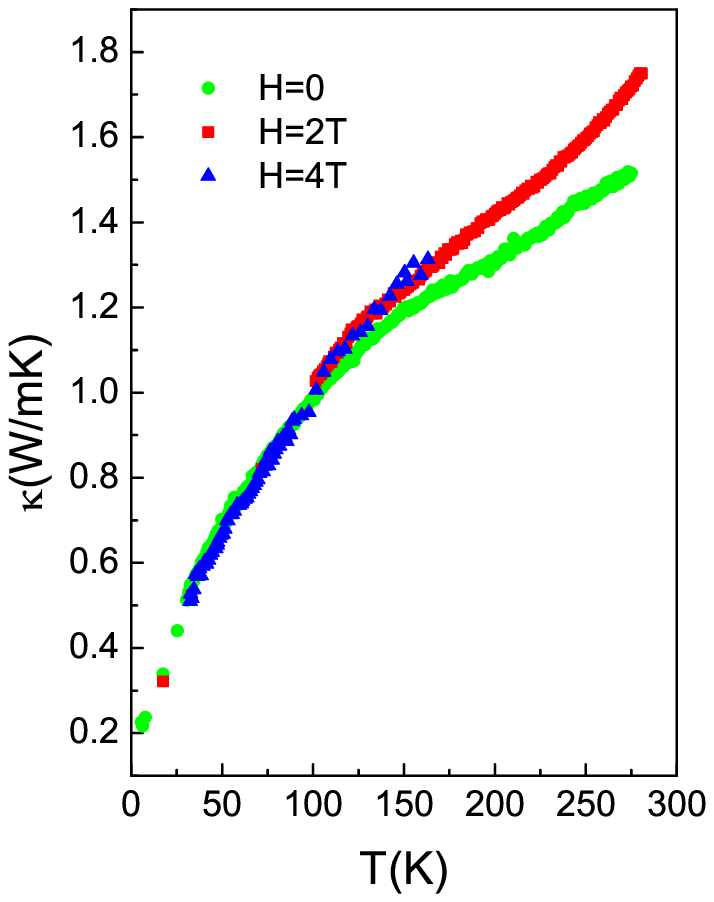}
\caption{\label{fig:5} (Color online) Temperature dependence of
thermal conductivity of polycrystalline $La_{0.8}Sr_{0.2}MnO_3$
for various magnetic fields.}
\end{figure*}
Samples were magnetically characterized by measuring the
temperature variation of magnetic AC susceptibility and DC
magnetization at different applied magnetic fields. The
temperature dependence of the in-phase component $\chi '$ of the
AC magnetic susceptibility reveals a typical ferromagnetic
behavior with a very smooth decrease of $\chi '$ below the maximum
located near $231 K$ (see Fig.~\ref{fig:2}(a)). The Curie
temperature $T_C$ (defined by the largest slope of $\chi '$) is
equal to $286 K$. The out-of-phase component $\chi ''$ of the AC
magnetic susceptibility (Fig.~\ref{fig:2}(b)) also exhibits a
maximum around $231 K$. It should be noted that an anomaly seen in
the out-of-phase component (b) just below $50K$ is simply an
artifact which arises from magnetic traces contained in a sample
holder. It is reproducible but it is not related to the true
manganite phase shown by the in-phase component (a). The
temperature dependence of the DC magnetization under $100$, $300$
and $500 Oe$ is shown in Fig.~\ref{fig:3}. For $H = 100$ and $300
Oe$ one may distinguish a pronounced split between the
field-cooled (FC) and zero-field-cooled (ZFC) curves appearing
below $230$ and $160 K$, respectively. For $H=500 Oe$ the FC and
ZFC curves practically coincide. Such a weak irreversibility
behavior suggests a rather low level of magnetic anisotropy in our
sample. The Curie temperature gradually decreases from $T_C =284
K$ (for $100 Oe$) to $T_C =281 K$ (for $500 Oe$).

The electrical resistivity was measured by the standard four-probe
method using currents up to $10 mA$ (see~\cite{8,9}). The
temperature dependence of the measured electrical resistivity
(Fig.~\ref{fig:4}) is typical for magnetoresistive materials,
exhibiting a two-phase behavior with a more conductive (metallic)
and a less conductive (insulating) regions separated by a
pronounced maximum at the peak temperature $T_p = 204 K$ (at zero
magnetic field). Unlike the previously discussed Curie temperature
variation, the resistance maximum gradually moves towards higher
temperatures (approximately proportionally to the applied magnetic
field) and reaches $T_p = 217 K$ at $H=8T$. This $T_p$ shift is
accompanied by a marked reduction of the resistance maximum
amplitude in the whole temperature range studied. The low
temperature values of the resistivity at 8T are reduced almost
three times as compared to the zero field case.

The thermal conductivity was measured using the stationary heat
flux method~\cite{10} in the temperature range $5-300K$. The
experimental setup and the measurement procedures have been
described in detail earlier~\cite{10,11}. The temperature gradient
along the sample was in the range $0.1-0.5K$. The magnetic field
was applied normally to the heat flow. Particular care was taken
to avoid a parasitic heat transfer between the sample and its
environment. The measurement error was below $2\%$ and the surplus
error (estimated from the scattering of the measurement points)
did not exceed $0.3\%$. Fig.~\ref{fig:5} depicts the obtained
temperature dependence of the thermal conductivity for various
magnetic fields.

\begin{figure*}
\includegraphics[width=7.0cm]{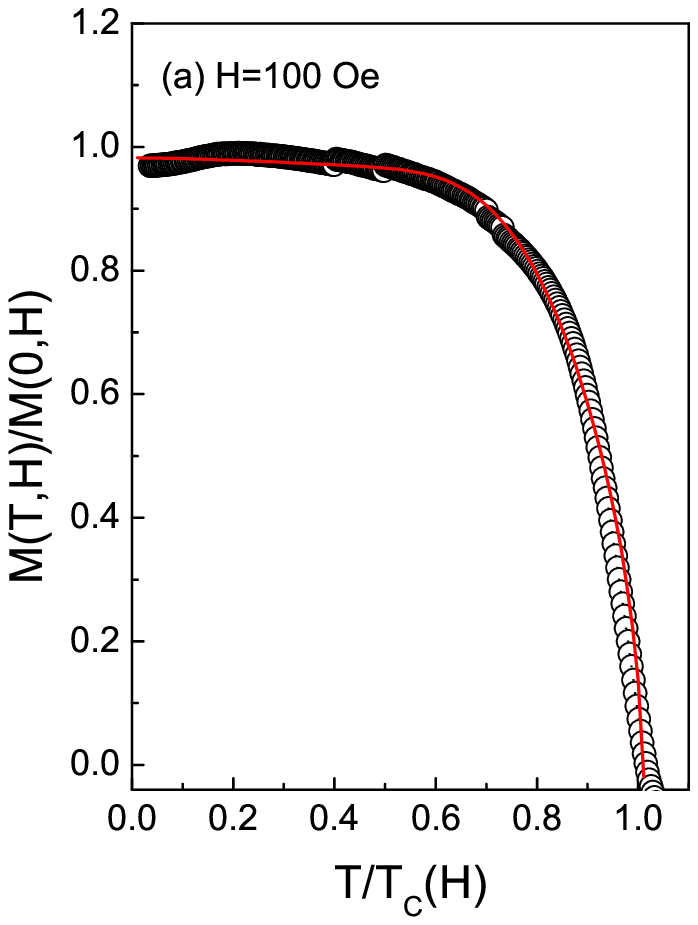}
\includegraphics[width=7.2cm]{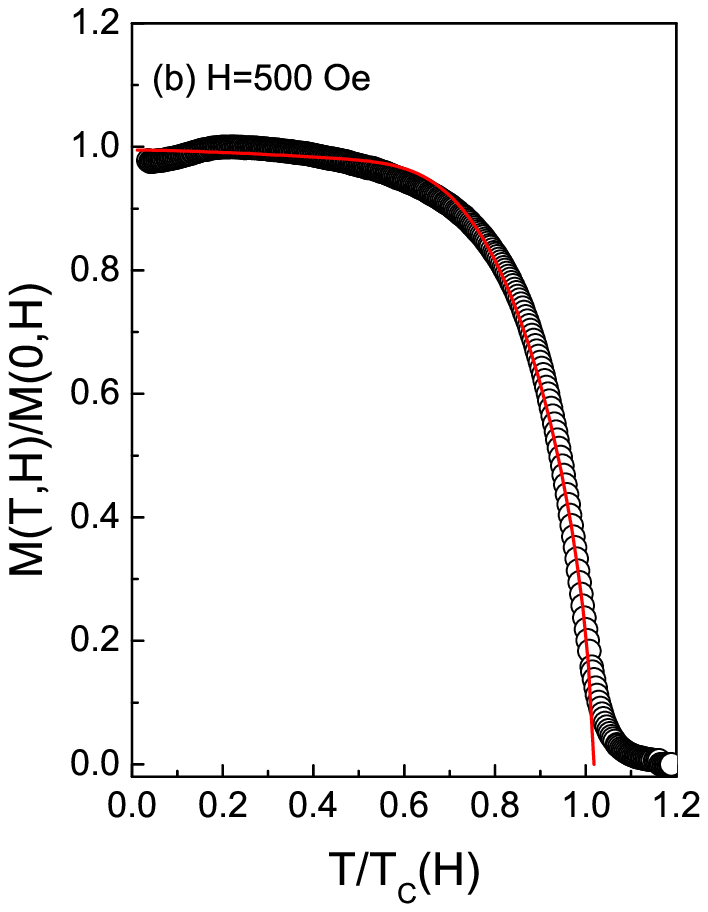}
\caption{\label{fig:6} (Color online)  The best fits (red solid
lines) of the temperature dependence of the normalized
magnetization $M(T,H)$ for (a) $H=100 Oe$ and (b) $H=500 Oe$,
according to Eq.(1).}
\end{figure*}

\section{III. Discussion}

\subsection{A. Magnetization}

Let us start the discussion of the obtained results with
thermodynamic properties and consider the behavior of a low-field
DC magnetization $M(T,H)$ in the FM region. Fig.~\ref{fig:6} shows
the best fits of the temperature dependence of the normalized
magnetization according to the following expression (which is an
analytical approximate solution of the Curie-Weiss mean-field
equations~\cite{12})
\begin{equation}
M(T,H)=M_{ext}(H)+M_0(H)\tanh\left\{\sqrt{\left[\frac{T_C(H)}{T}\right]^{2}-1}\right\}
\end{equation}
Here, $T_C(H)$ is the field-dependent Curie temperature,
$M_{ext}(H)$ is external contribution to magnetization in applied
field, and $M_0(H)$ accounts for deviation from the saturation
value of spontaneous magnetization $M_s$.  Using $T_C =286K$ and
$T_C =281K$ for the experimentally found values of the Curie
temperature for $H=100 Oe$ and $H=500 Oe$, respectively, the best
fits produced $M_0=0.9M_s$ and $M_{ext}=0.1M_s$  for the model
parameters.

Turning to the interpretation of the obtained results, notice that
within the whole temperature interval, the sample is practically
totally in a FM phase. Moreover, using a previously obtained
phenomenological formula~\cite{13} $T_{MI}=(1-4M_0/9M_s)T_C$ the
found deviation from the saturation magnetization $M_0=0.9M_s$
thereby allows for an estimate of the M-I transition temperature
$T_{MI}=0.6T_C$, which in turn provides to make a clear
thermodynamic distinction between the metallic ($T<T_{MI}$) and
semiconducting ($T>T_{MI}$) regions and complements the measured
value of the resistivity peak temperature $T_p$ (discussed in the
next Section).

\subsection{B. Resistivity}

\begin{figure*}
\includegraphics[width=7.2cm]{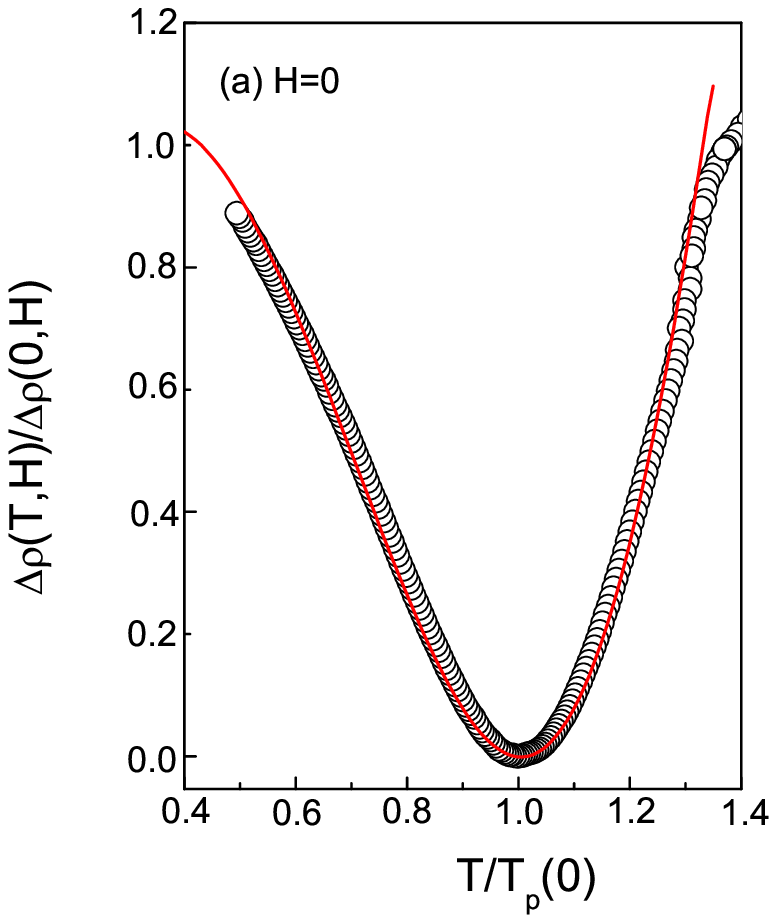}
\includegraphics[width=7.0cm]{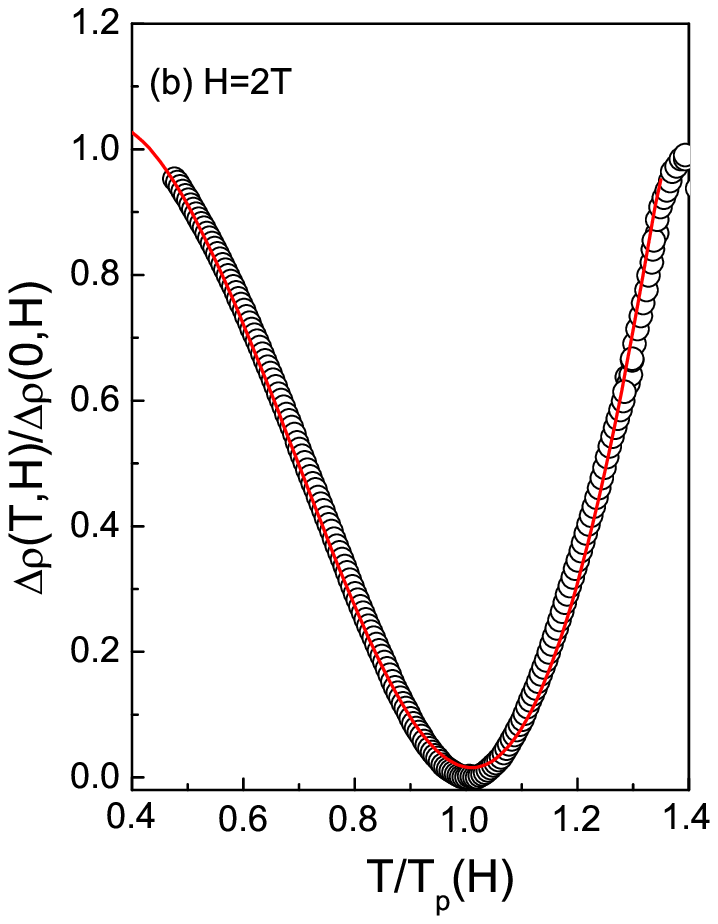}
\includegraphics[width=7.0cm]{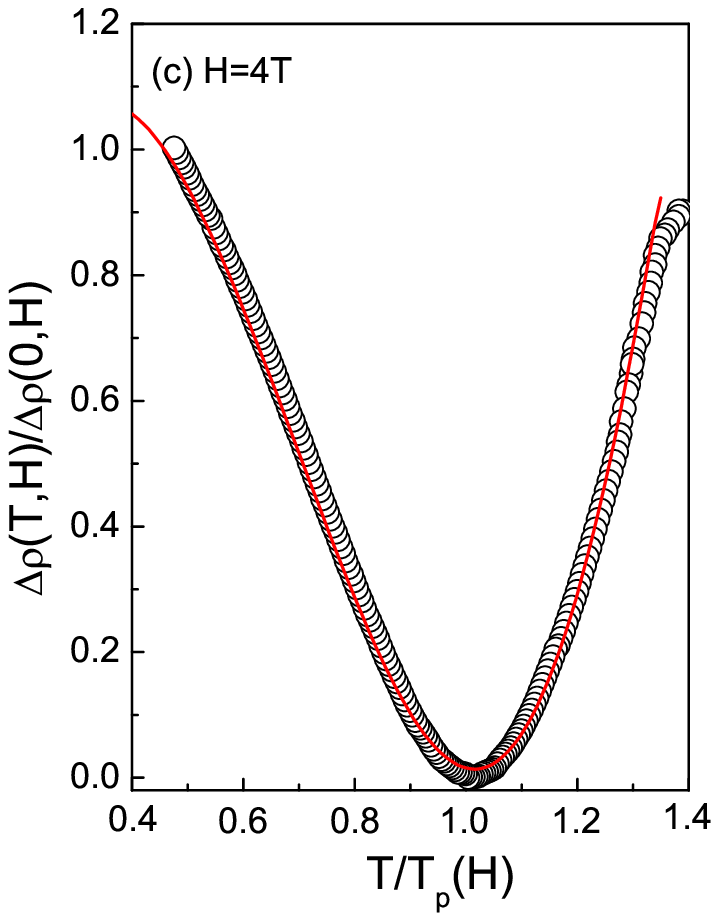}
\caption{\label{fig:7} (Color online) Temperature dependence of
the normalized resistivity $\Delta \rho (T,H)/\Delta \rho (0,H)$
versus the reduced temperature $T/T_p(H)$ for (a) $H=0$, (b)
$H=2T$, and (c) $H=4T$. The red solid lines are the best fits
according to the model equations (see the text).}
\end{figure*}

Turning to the discussion of the transport properties in our
polycrystalline samples, we notice that in order to exclude any
extrinsic effects (like grain boundary scattering), it is more
appropriate from a physical point of view to consider the
normalized resistivity $\Delta \rho (T,H)/\Delta \rho (0,H)$ where
$\Delta \rho (T,H)=\rho (T,H)- \rho (T_p,H)$ with $T_p$ being the
peak temperature and $\rho (0,H)$ the resistivity taken at the
lowest available temperature. Fig.~\ref{fig:7} depicts the
above-defined normalized resistivity versus the reduced
temperature $T/T_p(H)$ for (a) $H=0$, (b) $H=2T$, and (c) $H=4T$.
The solid lines are the best fits to theoretical laws for which we
outline the derivation in what follows. Let us start our
discussion with a brief outlook of the polaron hopping
conductivity scenarios. Recall that
several~\cite{12,13,14,15,16,17,18} approaches have been suggested
so far to tackle this problem. In essence, all of them are based
on a magnetic localization concept which relates the observable MR
at any temperature and/or applied magnetic field to the local
magnetization. In particular, one of the most advanced models of
this kind~\cite{15} ascribes the metal-insulator (M-I) like
transition to a modification of the spin-dependent potential $J_H
\vec s \cdot \vec S$  associated with the onset of magnetic order
at $T_C$ (where $J_H$ is the on-site Hund's-rule exchange coupling
of an $e_g$ electron with $s=1/2$ to the localized $Mn$ $t_{2g}$
ion core with $S=3/2$). Specifically, the hopping based
conductivity reads $\sigma (R)=\sigma _0(R)\exp [-U(R)]$ where
$U(R)=2R/L-W_{ij}/k_BT$ and $\sigma _0(R)=e^2R^2\nu _{ph}N_m$.
Here $R$ is the hopping distance (typically~\cite{12}, of the
order of $1.5$ unit cells), $L$ the charge carrier localization
length (typically~\cite{13,17}, $L=0.5nm$), $\nu _{ph}$ the phonon
frequency, $N_m$ the density of available states at the magnetic
energy $J_H$, and $W_{ij}$ the effective barrier between the
hopping sites $i$ and $j$. There are two possibilities to
introduce an explicit magnetization dependence into the above
model: either assuming a magnetization-dependent localization
length $L(M)$~\cite{16} which leads to an unusual thermal behavior
of the electrical resistivity~\cite{12,13,18} or through modifying
the hopping barrier assuming  $W_{ij}=W_{ij}(M)$~\cite{15}. The
second scenario results in a more conventionally acceptable
thermally activated behavior of MR over the whole temperature
range. Indeed, since a sphere of radius $R$ contains  $(4/3)\pi
R^{3}/v$ sites where $v=5.7\times 10^{-29}m^{3}$ is the lattice
volume per manganise ion, the smallest value of $W_{ij}$ is
therefore $[(4/3)\pi R^{3}N(E_m)]^{-1}$. Minimizing the hopping
rate, one finds that the conductivity should vary according to the
Mott law as follows
\begin{equation}
\sigma (T)=\sigma _0(T)\exp\left \{-\frac{T_p}{T}\left [1-\left
(\frac{M}{M_s}\right)^2\right]\right\}^{1/4}
\end{equation}
where
\begin{equation}
\sigma _0(T)=\sigma _m\sqrt{\frac{T_p}{T}}
\end{equation}
with
\begin{equation}
\sigma _m=\left (\frac{4}{9}\right )e^2L^2\nu _{ph}N_m
\end{equation}
and
\begin{equation}
T_p=\left (\frac{9}{4}\right )^3\frac{1}{2\pi k_BN_mL^3}
\end{equation}
Notice that this scenario was used by Wagner et al.~\cite{17} to
successfully interpret their MR data on low-conductive
$Nd_{0.52}Sr_{0.48}MnO_3$ films.
Fig.7(a) presents the best fit for zero-field temperature
dependence of the normalized resistivity $\Delta \rho (T,H)/\Delta
\rho (0,H)$ according to our Eqs.(1)-(3) assuming as usual that
$\rho =1/\sigma $. To make sure our scenario for spin polaron
dominated transport makes sense indeed, let us estimate the
"vital" model parameters. In particular, using the typical value
for the phonon frequency in this type of materials $\nu _{ph}
=2\times 10^{13}s^{-1}$, and the experimental value for $\sigma
_m$, our Eqs.(1)-(5) predict $L=0.5nm$ for the localization length
(or the size of the spin polaron) in good agreement with the
published values for this parameter~\cite{12,13,14,15,16,17,18}.
Besides, using this found $L$ and the experimental value of the
zero-field peak temperature $T_p =204K$, we get $N_m=2\times
10^{-28}m^{-3}eV^{-1}$ as a reasonable estimate of the carrier's
number density. Finally, using the experimental value of the
zero-field Curie temperature $T_C =286K$, the spin exchange value
of the coupling energy for our sample is estimated to be
$J_H=25meV$ as is expected for this parameter~\cite{19}.

Turning to the influence of field effects on the temperature
behavior of the resistivity in our sample, it is important to
notice the different evolution of the two characteristic
temperatures with the applied field $H$ which results in a marked
modification of the boundary between the more conducting and less
conducting regions. Indeed, while the peak temperature increases
with $H$ reaching the value of $T_p=210K$ at $H=2T$, the Curie
temperature, on the other hand, goes in the opposite direction,
strongly decreasing from $T_C =286K$ at $H=0$ to $T_C =260K$ at
$H=2T$. At the same time, a thorough analysis of the high-field
results for the magneto-resistivity revealed that a simple
modification of the hopping expression given by Eqs.(1)-(3) to
non-zero magnetic fields is not enough to fit our MR data and some
extra contribution is needed. Eventually, we found that our data
for $H=2T$ and $H=4T$ shown in Fig.7(b) and Fig.7(c) can be quite
successfully fitted assuming an appropriate field dependence of
both critical temperatures, $T_C(H)$ and $T_p(H)$, and using the
following expression
\begin{equation}
\sigma (T,H)=\sigma _1(T,H)+\sigma _2(T,H)
\end{equation}
where $\sigma _1(T,H)$ is simply the field-induced hopping
conductivity generalizing Eq.(2), that is
\begin{equation}
\sigma _1(T,H)=\sigma _{01}(T,H)\exp\left [-U_1(T,H)\right ]
\end{equation}
with
\begin{equation}
\sigma _{01}(T,H)=\sigma _{m1}\sqrt{\frac{T_p(H)}{T}}
\end{equation}
and
\begin{equation}
U_1(T,H)=\left \{-\frac{T_p(H)}{T}\left [1-\left
(\frac{M}{M_s}\right)^2\right]\right\}^{1/4}
\end{equation}
Notice that the temperature and field dependence of DC
magnetization $M(T,H)$ in the above equations is still governed by
the universal expression given by Eq.(1). On the other hand,
assuming a linear superposition of effects as in any linear
response theory we propose the following empirical expression for
an extra contribution to the magneto-conductivity
\begin{equation}
\sigma _2(T,H)=\sigma
_{02}(T,H)[pe^{-U_2(T,H)}+(1-p)e^{+U_2(T,H)}]
\end{equation}
where
\begin{equation}
\sigma _{02}(T,H)=\sigma _{m2}\left [\frac{T}{T_0(H)}\right
]^{1/4}
\end{equation}
and
\begin{equation}
U_2(T,H)=\frac{T}{T_0(H)}
\end{equation}
Let us consider the origin of the $\sigma _2(T,H)$ term in the
total MR in our material. As a matter of fact, the very form of
Eq.(10) suggests that this contribution describes equilibrium
state of a two-level system (with fractional population $p$)
created by spin-dependent energy splitting in applied magnetic
field~\cite{20}. More specifically, this contribution is governed
by a Zeeman like term $W_H=-\mu (R)H$ where $\mu (R)=\pi R^2gS\mu
_BL/v$ is the local magnetic moment of spin polaron~\cite{21}.
Here, $R$ is the hopping distance, $v$ is the lattice volume per
manganite ion, $\mu _B$ is the Bohr magneton and $g$ is the
gyromagnetic ratio. It should be noted that unlike the previous
contribution, given by the $\sigma _1(T,H)$ term, the second
contribution does not exhibit a thermally-activated behavior even
though (as we shall see) it is still related to the adopted in
this paper spin polaron hopping scenario.

Furthermore, to correctly describe the semiconducting region
(above $T_{MI}=0.77T_C$), it is important to take into account
thermally excited polaron states~\cite{22} with the number density
$n(T)=(2\pi m_pk_BT/\hbar ^2)^{3/2}$ where $m_p$ is an effective
polaron mass.

Based on the above assumptions, the second contribution to the
transport mechanism in our material can be presented in the
general form of a field-dependent hopping conductivity as follows:
\begin{equation}
\sigma _2(R,H)=\sigma _{02}(R)[pe^{-U_2(R,H)}+(1-p)e^{+U_2(R,H)}]
\end{equation}
where
\begin{equation}
U_2(R,H)=\frac{2R}{L}-\frac{\mu (R)H}{k_BT}
\end{equation}
and
\begin{equation}
\sigma _{02}(R)=e^2R^2\nu _{ph}N_s(R,T)
\end{equation}
Here $N_s(R,T)=\sqrt{n(T)/N(R)J_H^2v}$ is the local number density
of polaron states (including thermally excited carriers in the
semiconducting region) with $N(R)=3\pi R^3/4v$ being the number of
available sites.

Minimizing the hopping rate given by Eqs. (13)-(15), we find that
the second contribution to the MR is indeed governed by
Eqs.(10)-(12) with
\begin{equation}
\sigma _{m2}(H)=e^2L^2\nu _{ph}\left (\frac{2m_p}{h^2L^2}\right
)^{3/4}\left [\frac{36\pi ^2}{k_BT_0(H)}\right ]^{1/4}
\end{equation}
and
\begin{equation}
T_0(H)=\frac{4\pi L^3gS\mu _BH}{k_Bv}
\end{equation}
Using $T_C =260K$ and $T_p =210K$ for the experimentally found
values of the Curie temperature and the resistivity peak
temperature at $H=2T$, respectively, the best fits were obtained
for the following set of the model parameters:  $T_0 =240K$,
$M_{ext}=0.47M_s$, $M_0=0.53M_s$, and $p=0.72$. Furthermore,
since~\cite{13} $T_{MI} =(1-4M_0/9M_s)T_C$ the above deviation
from the saturation magnetization gives $T_{MI}=0.77T_C$ for an
estimate of the M-I transition temperature at $H=2T$ (to be
compared with $T_{MI}=0.6T_C$ at $H=0$). Along with a similar
behavior of the peak temperature $T_p$, this implies an effective
extension of the more conductive phase to higher temperatures (in
contrast with the field-free case shown in Fig.7(a)). Using the
previously discussed values of the phonon frequency $\nu _{ph}$,
the size of the spin polaron (localization length) $L$, and the
experimental values for $\sigma _{m1}$ and $\sigma _{m2}$, from
Eqs. (16) and (17) we obtain $m_p=3.2m_e$ for a reasonable
estimate of the effective polaron mass~\cite{2}. Furthermore, let
us estimate the absolute values of the Curie temperature $T_C(H)$
and the electron temperature $T_{MI}(H)$ for $H=2T$. Making a
reasonable assumption that $k_BT_C(H)=k_BT_C(0)-\mu (L)H$ with
$k_BT_C(0)=J_H=25meV$ being a zero-field value and $\mu (R)$ being
defined earlier, we find $k_BT_C(H)=20meV$ or $T_C(H)=260K$, in
agreement with the observations. Likewise, the field dependence of
$T_{MI}(H)$ is governed by the corresponding behavior of the
density of polaron states as follows $T_{MI}(H)\propto
1/N_m[J_H-\mu (L)H]$. Since $\mu (L)H/J_H \ll 1$, we find that
$T_{MI}(H)\simeq T_{MI}(0)[1+\mu (L)H/J_H]$ where
$T_{MI}(0)\propto 1/N_m(J_H)$, so that $T_{MI}(H)=200K$.

Finally, it is worth commenting on the value of the fractional
population $p$.  In magnetically disordered well-defined
paramagnetic phase one would expect equal distribution of
populations of spin polarons with $p=1/2$.  The fact that our
experiments instead predict $p=0.72$ suggests rather strong
field-induced polarization effects at $H=2T$ in our material,
indicating the presence of ordered FM regions in the
semiconducting phase as well.

\subsection{C. Thermal conductivity}

\begin{figure*}
\includegraphics[width=7.0cm]{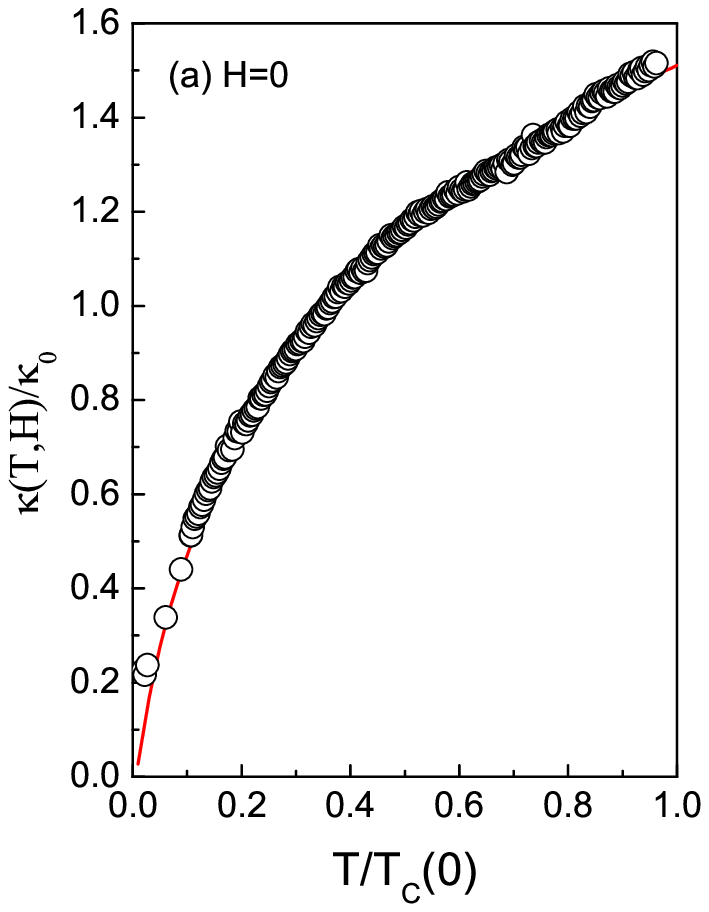}
\includegraphics[width=7.0cm]{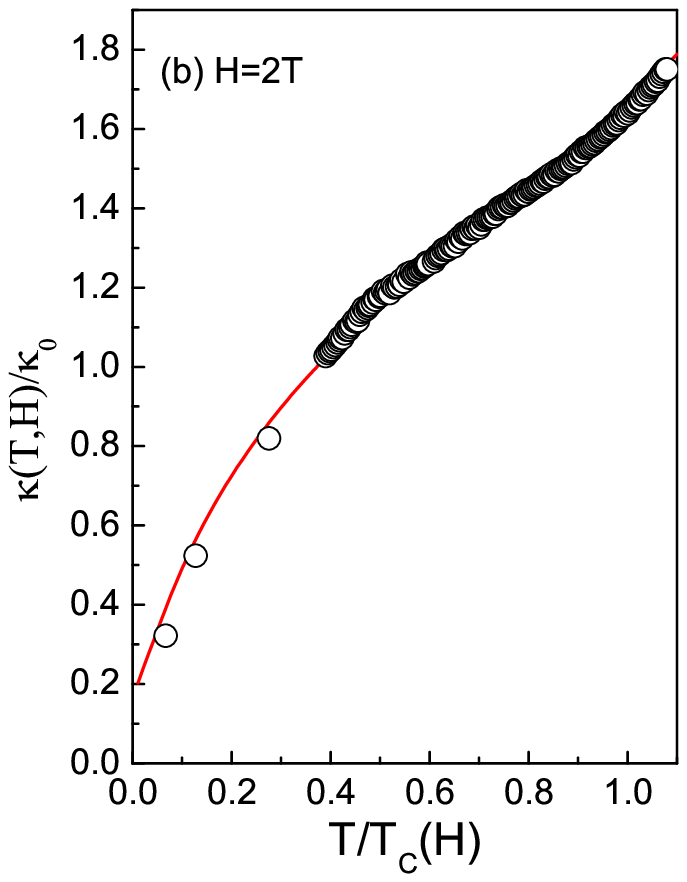}
\includegraphics[width=7.0cm]{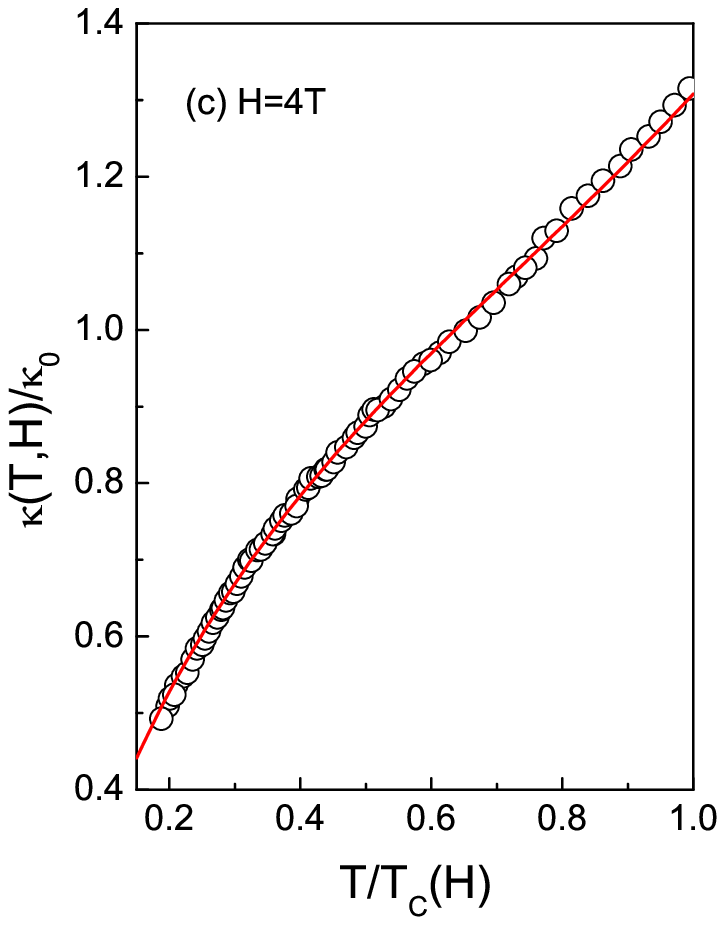}
\caption{\label{fig:8}(Color online) Temperature dependence of the
thermal conductivity $\kappa (T,H)/\kappa _0$ versus the reduced
temperature $T/T_C(H)$ for (a) $H=0$, (b) $H=2T$, and (c) $H=4T$.
The red solid lines are the best fits according to the model
equations (see the text).}
\end{figure*}

Based on the above experimental background and our previous
experience in studying  thermoelectric response of similar class
of materials~\cite{13}, it is quite reasonable to assume that the
heat transport properties of our sample are dominated by hopping
(tunneling) of spin polarons and a Zeeman term as well. When
discussing the relationship between electric and heat transport
behavior in both metallic and non-metallic systems, the main
question which always needs to be addressed (and still remains
controversial) is about the validity of the so-called
Wiedemann-Franz (WF) law. For example, a recent report by Lia et
al.~\cite{14} suggests that the ratio of the thermal conductivity
to the electrical conductivity in their $Nd_{0.75}Na_{0.25}MnO_3$
samples strongly deviates from the WF law even in the FM metallic
state. Our present study, on the other hand, seems to be in total
support of the WF law for the whole temperature interval because,
based on the above assumptions, we are able to successfully fit
our zero-field data for the thermal conductivity shown in
Fig.~\ref{fig:8}(a) using the following expression (Cf. with
similar Eqs.(2)-(9) for the resistivity)
\begin{equation}
\kappa (T)=\kappa _{0}(T)\exp\left [-U(T)\right ]
\end{equation}
where
\begin{equation}
\kappa _{0}(T)=\kappa _{m}\sqrt{\frac{T}{T_p}}
\end{equation}
and
\begin{equation}
U(T)=\left \{-\frac{T_p}{T}\left [1-\left
(\frac{M}{M_s}\right)^2\right]\right\}^{1/4}
\end{equation}
The direct comparison with Eq.(3) reveals that our fitting
expression for the pre-factor $\kappa _{0}(T)$ given by Eq.(19) is
nothing else but a manifestation of the WF law relating the
electric and thermal conductivities because
\begin{equation}
\kappa _{0}(T)=\frac{\pi ^2k_B^2T}{3e^2}\sigma _{0}(T)=\kappa
_{m}\sqrt{\frac{T}{T_p}}
\end{equation}
with
\begin{equation}
\kappa _{m}=\frac{27\pi k_B\nu _{ph}}{32L}
\end{equation}
Likewise, high-field results for the thermal conductivity are
found to be well described  using the field-induced modification
of hopping scenario assuming the validity of the WF law. Namely,
Fig.~\ref{fig:8}(b) and Fig.~\ref{fig:8}(c) show the best fits for
$H=2T$ and $H=4T$ using the following expression which takes into
account both hopping ($\kappa _1$) and  Zeeman ($\kappa _2$) terms
\begin{equation}
\kappa (T,H)=\kappa _1(T,H)+\kappa _2(T,H)
\end{equation}
where
\begin{equation}
\kappa _1(T,H)=\kappa _{01}(T,H)\exp\left [-U_1(T,H)\right ]
\end{equation}
with
\begin{equation}
\kappa _{01}(T,H)=\frac{\pi ^2k_B^2T}{3e^2}\sigma
_{01}(T,H)=\kappa _{m1}\sqrt{\frac{T}{T_p(H)}}
\end{equation}
while a Zeeman-type two-level induced contribution reads
(similarly to Eq.(10))
\begin{equation}
\kappa _2(T,H)=\kappa
_{02}(T,H)[pe^{-U_2(T,H)}+(1-p)e^{+U_2(T,H)}]
\end{equation}
where
\begin{equation}
\kappa _{02}(T,H)=\frac{\pi ^2k_B^2T}{3e^2}\sigma
_{02}(T,H)=\kappa _{m2}\left [\frac{T}{T_0(H)}\right ]^{1/4}
\end{equation}
with
\begin{equation}
\kappa _{m2}(H)=\sqrt{\frac{\pi ^2}{12}}\left (\frac{k_B\nu
_{ph}}{L}\right )\left [\frac{2m_pk_BT_0(H)L^2}{\hbar ^2}\right
]^{3/4}
\end{equation}
and with $T_0(H)$ still given by Eq.(17). It is important to note
that the best fits for the magneto-thermal conductivity were
obtained for the set of model parameters used in the previous
Section for fitting of our magneto-resistivity data. In
particular, the $H=2T$ data were fitted using:  $T_C =260K$, $T_p
=210K$, $T_0 =240K$, $M_{ext}=0.47M_s$, $M_0=0.53M_s$, and
$p=0.72$.

\section{IV. Conclusions}

It was pointed out~\cite{23,24} on investigations of
magnetotransport in charge ordered manganites with similar
magnetic ground states that the origin of magnetoresistance cannot
be concluded from the isofield resistivity measurements alone.
Hence it is of importance to organize experimental runs on related
properties (equilibrium and nonequilibrium ones) quasi
simultaneously on the same materials. Moreover it was found that
neutron spectra of CMR materials cannot be usually explained with
a picture of spin clusters moving in a paramagnetic background
(i.e. magnetic polarons). Rather, a model describing an average
magnetic coherence extending over several $Mn$ spins leads to
better fits to the data (see Viret et al.~\cite{4}).  That is why
it is interesting to examine a possible interplay between
different magnetic states including both localized and collective
excitations. In Ref.~\cite{3}, some regimes had already been
pointed out through similarities (and differences) between the
specific heat data and the electrical resistivity. In order to
further shed some light on a possible connection between spin and
charge degrees of freedom in CMR exhibiting manganites, a thorough
experimental study of two magnetic and two transport properties of
magneto-resistive polycrystalline $La_{0.8}Sr_{0.2}MnO_3$ samples
has been presented in this paper. Using a properly generalized
analytical expression for the spontaneous DC magnetization
[Eq.(1)], the temperature and magnetic field dependencies of
electrical resistivity [Eq.(6)] and thermal conductivity [Eq.(23)]
were successfully fitted, assuming spin polaron hopping scenario
(strongly influenced by a Zeeman type splitting effects), presence
of thermally excited polaron states (needed to correctly describe
the semiconducting region), and the validity of the
Wiedemann-Franz law. It is important to underline the
self-consistency and coherence of our approach which produced very
reasonable estimates for numerical values of microscopic
parameters. It is also worth noting that the presented
experimental and theoretical results corroborate magneto-transport
measurements and subsequent analysis on the temperature behavior
of previously obtained magneto-thermopower results on a similar
material~\cite{13}. And finally, a brief comment is in order on
the role of grain-boundary effects in the transport properties
under discussion. The very fact that the adopted here polaron
picture reasonably well describes {\it both} electric resistivity
and thermal conductivity suggests a rather high quality of our
sample (which is also evident from its X-ray diagram shown in
Fig.1) with presumably narrow enough grain distribution and
quasi-homogeneous low-energy barriers between the adjacent grains.
Besides, in order to minimize the inevitable influence of the
grain-boundary scattering effects on the metal-insulator
transition temperature in polycrystalline samples (see,
e.g.,~\cite{25} for detailed discussion), the latter is defined
via the properly {\it normalized} resistivity data (see
Fig.~\ref{fig:7}).

{\it Note added}: when our paper was completed, we became aware of
the very important experimental results on {\it direct}
observation of polaron states and nanometer-scale phase separation
in CMR exhibiting manganites. More specifically, using high
resolution topographic images obtained by scanning tunneling
microscope, regular stripe-like or zigzag patterns on a width
scale ranging between $0.4nm$ and $0.5nm$ were observed~\cite{26}
in the insulating state of $Pr_{0.68}Pb_{0.32}MnO_3$ which
remarkably correlate with the size of spin polarons ($L=0.5nm$)
deduced from our present measurements on $La_{0.8}Sr_{0.2}MnO_3$
providing thus further evidence in support of our interpretation
based on spin polaron scenario.

\section{acknowledgments}

This work was partially supported by Brazilian agency CAPES,
Ministry of Science and Higher Education (PL) - grant
Walonia/286/2006, Kasa Mianowskiego (PL), CGRI (B) and FNRS (B) in
Liege through B 8/5-CB/SP-9.154 and FRFC 1.5.115.03 convention.

\end{document}